# OBJECT ORIENTED MODELLING OF IDEA USING GA BASED EFFICIENT KEY GENERATION FOR E-GOVERNANCE SECURITY (OOMIG)

Arindam Sarkar [1], S. Karforma [2], J. K. Mandal [3]

[1]Department of Computer Science & Engineering, University of Kalyani, W.B, India
arindam.vb@gmail.com
[2]Department of Computer Science, The University of Burdwan, W.B, India
dr.sunilkarforma@gmail.com
[3]Department of Computer Science & Engineering, University of Kalyani, W.B, India
jkm.cse@gmail.com

## ABSTRACT

*Nowadays different state and central government in India as well as abroad are taking initiative to deliver different kind of services electronically specially using Information and Communication Technology (ICT). Intruders or hackers can steal or modify the information communicated between Government and consumer through internet. To implement privacy and confidentiality of the information we must use suitable encryption technique. In this paper an Object Oriented Modelling of International Data Encryption Algorithm (IDEA) using GA based efficient key generation technique has been proposed to incorporate privacy and confidentiality of information which would be communicated between government and consumer.*

## 1. INTRODUCTION

In early nineties India started implementing the uses of Information and Communication Technology i.e ICT in various fields to deliver different services efficiently, effectively and securely among its citizens [1]. This period witnessed the silent revolution in the domain of (ICT) and this actually accelerates the propulsion to reach and cater to the basic needs of the citizenry residing in the remote fringes of this country, having limited access to basic civic amenities. The authors are of the view that country's one billion plus population with wide divergences in its demographic pattern really enhances the complexity of the governance of this country. Catering public services to this huge population is a great challenge to any civil administration. E-governance [2] can be a suitable solution to this problem. The primary delivery models of e-Government can be divided into Government-to-Citizen or Government-to-Consumer (G2C), Government-to-Business (G2B), Government-to-Government (G2G), Government-to-Employees (G2E). Authors are of the opinion that though Indians have a strong presence in the software world, digitization of governmental services is still at its infancy in India. The present study tries to impose security in e-Governance transaction aspects where the implementation of e-Governance and the Information and Communication Technology (ICT) [3 , 4, 5, 6, 7] can be used to solve  various needs of the masses. While accessing public services, citizens need to possess multiple instruments which will only validate their accessibility. Present instruments are very much requirement specific, but the authors are of the view that there should be only one electronic card with multipurpose functionality because issuance of multiple cards to an individual is not only economically infeasible but also leads to anomalies in data integrity [8]. This Multipurpose Electronic Card (MEC) will be generated by the Government for its citizenry. For a new born baby the instrument will be issued from the lowest level of governance which will contain the information of an individual in encrypted form. Here involvement of cryptography becomes very much relevant in order to achieve the highest level of security.

Cryptography is the science and art of encryption and decryption of a message. In secret key cryptography same key is used for encryption and decryption of message. Encryption and decryption when done with two different keys, it is called Public key algorithm. Secret key cryptographic algorithms which operate on a plaintext a single bit at a time, it is called Stream cipher. Secret key algorithm which operates on a plaintext in groups of bits (i.e blocks) is called Block cipher. International Data Encryption Algorithm (IDEA) [9, 10] is an industry standard block cipher, one of the most secure algorithms in its class. Secret key algorithm can be explained in manner of M= Message (or plain text), C= Cipher Text, E= Encryption function, D= Decryption function, K= Encryption & Decryption key, Encryption denotes: $E_K(M)=C$ , Decryption denotes: $D_K(C)=M$. From this expressions, we can say, $D_K [ E_K (M) ] =$ M. Individual have to submit this instrument to avail the allocated facilities. Facilities will be authenticated by checking the status after decrypting the unique id. Facilities will be updated with due course of time and simultaneously with the up gradation of the individual. To implement e-Governance, Union Government approved the National e-Governance Plan (NeGP), comprising of 27 Mission Mode Projects (MMPs) and 8 components, on 2006. The ongoing projects in the MMP category, being implement by various Central Ministries/State departments/States would be suitably augmented to align with the objectives of NeGP [3].The Mission Mode Projects and it's as envisaged in the NeGP is tabulated in table II. NeGP is the backbone for any e-Governance initiative that has been operationalised.

The authors are of the view that this Multipurpose Electronic Card (MEC) will be successfully implemented if IDEA algorithm is wrapped in Object Oriented Modeling. Object oriented modeling is used to demonstrate the following prospective in descriptive form. In case of software development we can use Object Oriented Programming instead of Procedure Oriented Programming. By using OOP implementation of IDEA, we have used class, instance, method, inheritance, data security features of OOP successfully. These are described as follows

***Class -*** A Class is template for an object, a user-defined data type that contains the variables, properties and methods in it. A class defines the abstract characteristics of a thing (object), including its characteristics (its attributes, fields or properties) and the thing's behaviors (the things it can do, or methods, operations or features). In our proposed model we have used four classes namely, CommonFields, SecurityInterface, Sender and Receiver. The necessary information's of an individual are the attributes of the class CommonFields. Besides that, this class contains various member methods to perform storage and retrieval of these attributes. Class Security Interface does not contain any attributes of its own as it inherits the class CommonFields in public mode. This class contains various member methods among which encryption and decryption methods are mainly used to implement the IDEA algorithm. Class sender inherits class SecurityInterface into public mode and contains member method to grant various facilities to the consumer from the Government end. A class receiver inherits the class SecurityInterface into public mode and contains member method of its own to check the privileges allocated to the consumer.

***Instance -*** One can have an instance of a class; the instance is the actual object created at run-time. In this paper the instance of class SecurityInterface is created in the main method which is used to call the various methods to implement the IDEA algorithm with OOP concept.

***Method -*** Method is a set of procedural statements for achieving the desired results. It performs different kinds of operations on different data types. This Multipurpose Electronic Card (MEC) is implemented based on various methods namely getbasicdata(), showbasicdata(), getmsg(), encryption(), decryption(), etc. Rest of the names of member methods are written in details in the class diagram.

***Inheritance -*** Inheritance is a process in which a class inherits all the state and behaviour of another class. This type of relationship is called child-Parent or is-a relationship. "Subclasses" are more specialized versions of a class, which inherit attributes and behaviours from their parent classes, and can introduce their own. In this article, class SecurityInterface inherits the class CommonFields in public mode. Similarly, class Sender and class Receiver inherits the class SecurityInterface in public mode.

***Data hiding –*** The prime objective of this paper is to incorporate the data security feature of IDEA algorithm with the object oriented programming so as to design a robust mechanism. And to achieve this property, authors declared all the data members of the class CommonFields as private which can only be accessed by the member methods of the same class and not by any other methods.

The organization of this paper is as follows. Section II deals with related works .Section III of the paper deals with architecture of G2C model with relative diagram**.** Proposed 128 bit key generation scheme for IDEA using GA algorithm has been discussed in section IV. Section V is deals with the proposed object oriented modelling of IDEA algorithm with the help of class diagram and sequence diagram. Conclusions & future scopes are drawn in section VI and that of references at end.

## 2. RELATED WORK

Security in E-Governance is already addressed by several researchers. Public key and private key cryptography are also used in E- Governance for implementing security. Some research papers proposed a encryption technique to implement confidentiality of information in G2G model employing object oriented concepts, wrapping DES algorithm and implementing confidentially of the message.

## 3. ARCHITECTURE OF A SECURED GOVERNMENT 2 CONSUMER MODEL

The G2C services of e-Governance can be further discussed as follows:

Government-to-Citizen [11] (abbreviated G2C) is the communication link between a government and private individuals or residents. Such G2C communication most often refers to that which takes place through Information Communication Technologies (or ICTs), but can also include direct mail and media campaigns. G2C can take place at the federal, state, and local levels. G2C stands in contrast to G2B, or Government-to-Business networks. One such Federal G2C network is USA.gov: the United States' official web portal, though there are many other examples from governments around the world [12].

**Architecture of Government 2 Consumer model**

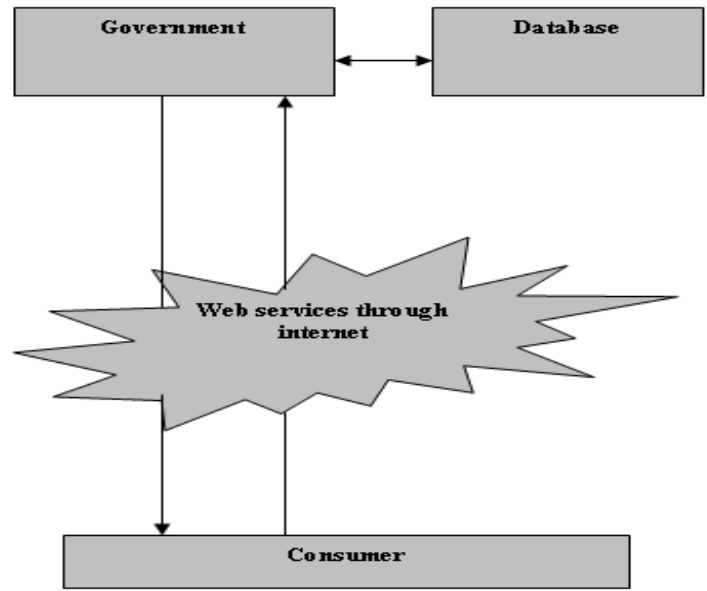

Figure 1.        Architecture of Government 2 Consumer Model

Generally the G2C model of e-Governance caters the needs of the citizens through web portals. A web portal displays various facilities available for the consumers. Consumers browse these web portals for claiming these services. Government stores the necessary information of the consumer to the database which is attached with the G2C (e.g www.incometaxindia.gov.in) portals and grant the facilities to its citizens through proper procedure. In our proposed G2C model the Government will deliver all its information in encrypted form to consumer and the consumer will decrypt it to reveal the original message. Government and consumer both will share the same key for encryption and decryption respectively.

## 3. PROPOSED 128 BIT KEY GENERATION SCHEME FOR IDEA USING GA

The technique considers the generation of purely random bits in the form of blocks of bits with 128 bits in length becomes the 1st chromosome of the 1st generation. Randomization is made using (eq 1) along with genetic function such as crossover & mutation. Chromosomes are represented as binary string with population size 10 [13, 14, 17, 18, 19, 20]

$$X_{n+1} = (a * X_{n+c}) \bmod m \qquad (1)$$

Where, **Xn** denotes the randomly selected 128 bits which becomes first chromosome of the first generation, **m** -modulus (m > 0), a - multiplier (0 <= a < m), **c** -increment (0 <= c < m).

After crating the 1st generation of the chromosomes, next generation chromosomes are generated using the GA operators crossover followed by mutation.

If generation number is even, crossover is taken place from top to bottom in a sequential way by pairing up chromosomes like (1,2) (3,4) (5,6).In case of odd generation crossover is take place from bottom to top in a sequential way by pairing up chromosomes like (50,49) (48,47) (46,45).Consider 284, 7000, 9024, 4025, 1235, 2564, 654, 6526, 3652, 124, as first generation chromosomes.

2nd generation chromosomes are obtained by Pairing up like (284, 7000), (9024, 4025), (1235, 2564), (654, 6526), (3652, 124). For this generation crossover and mutation will take place let at 4th locus of the gene of chromosome.

Consider binary representation of chromosomes as 284= 0000100011100 7000= 1101101011000

The output stream on crossover is 0000**101011000**     1101**100011100**.

The output generated on mutation is     00011010101000=856 and 1100100011100=6428 (Mutation of 4th bit). Up to an N generation same technique is use to generate N*10 chromosomes (Each generation 10 populations) and get the final set of numbers.

The following steps are performed to encrypt using 128 bit keys.

1. Once all the numbers are generated then let this array of numbers be called GENETIC_ARRAY and CALCULATED the SUMMATION OF EACH DIGIT of a number until it becomes ONE DIGIT TERM (2365→ 2+3+6+5=16 → 1+6=7). These new collection of numbers is generated and let this collection is called CODED_ARRAY.

2. Use this numbers from CODED _ARRAY sequentially for substituting on a one-to-one basis for the characters of the PASSWORD text.

3. Use ASCII values of the PASSWORD (minimum 8 characters) characters and ADD the numbers of CODED_ARRAY from the ASCII values. For example the PASSWORD, the CODED TEXT will be calculated according to following method.
LET GENETIC _ARRAY = {2165, 1000, 8526, 7198, 4142, 4684, 2981, 9472}.
So CODED_ARRAY = {5, 1, 3, 5, 2, 4, 2, 4}.

Table 1 shows the results of the process.

Table 1. Genetic Algorithm Based Encryption

| Password Character | ASCII Value | GENETIC_ARRAY Number Taken sequentially | Addition | Result |
|---|---|---|---|---|
| P | 80 | 5 | 80+5 | 85(W) |
| A | 65 | 1 | 65+1 | 66(B) |
| S | 83 | 3 | 83+3 | 86(X) |
| S | 83 | 5 | 83+5 | 88(Z) |
| W | 85 | 2 | 85+2 | 87(Y) |
| O | 79 | 4 | 79+4 | 83(U) |
| R | 82 | 2 | 82+2 | 84(V) |
| D | 68 | 4 | 68+4 | 72(H) |

From table1 it is seem that the intermediate CODED text is "WBXZYUVH". Now this CODED text is converted into binary bits. 8 bits are taken from MSB position and divided into two equal parts. After obtaining the decimal equivalent of the 4 bit number, determine whether the value is even or odd. If it is even then 0 is stored as code value, else 1 is stored as code value. For the other 4 bit blocks, the same operation is performed. Now, the decimal value of the 4 bit number is represented according to its rank in the even or odd number series (considering 0 as the 1st even number and 1 as the first odd number). The rank is represented in binary number by maximum bits to represent a rank of a 4 bit even or odd number. A series of '0' (maximum 7 in number) should be inserted as padding bits to make it a simple multiple of 8. As a result of the operation, a 16 bit number is generated. If the     process is repeated for the other 56 bits by taking 8 bits from them, then the 64 bit number will become 128 bit number. The 128 bit number will act as the Key for IDEA encryption algorithm. The process for converting 64bit number to 128 bit number is termed as "key expansion method".

# 4. IDEA ALGORITHM

The following algorithm demonstrates the sequence of flow of control during transmission of encrypted data from government end [15].

## 4.1 Encryption Algorithm

***Step1***: Input the message which is 64 bit, key is 128 bit.
***Step2***: Break the 64 bit message into four 16 bit sub-block, x1, x2, x3, x4.
***Step3***: Break the key into six 16 bit sub keys z1, z2, z3, z4, z5, z6.
***Step4***: Multiply x1 and the first sub key.
***Step5***: Add x2 and the second sub key.
***Step6***: Add x3 and the third sub key.
***Step7***: Multiply x4 and the fourth sub key.
***Step8***: XOR the results of steps (4) and (6).
***Step9***: XOR the results of steps (5) and (7).
***Step10***: Multiply the results of step (8) with the fifth sub key.
***Step11***: Add the results of steps (9) and (10).
***Step12***: Multiply the results of step (11) with the sixth sub key.
***Step13***: Add the results of steps (10) and (12).
***Step14***: XOR the results of steps (4) and (12).
***Step15***: XOR the results of steps (6) and (12).
***Step16***: XOR the results of steps (5) and (13).
***Step17***: XOR the results of steps (7) and (13).
***Step18***: Derive four sub-blocks that are the output of steps (14), (15), (16), and (17). nterchange the two inner blocks, except for the last round, and input it to the next round.
***Step19***: After eighth round, start final output transformation.
***Step20***: Multiply x1 and the first sub key.
***Step21***: Add x2 and the second sub key.
***Step22***: Add x3 and the third sub key.
***Step23***: Multiply x4 and the fourth sub key
***Step24***: Group four sub-blocks to generate the cipher text.
***Step25***: End

## 4.2 Decryption Algorithm

The algorithm demonstrates the sequence of flow of control during decryption part of IDEA

***Step1***: Start
***Step2***: Accept unique ID received from Government.
***Step3***: Divide 128-bit key into eight 16-bit sub    keys
***Step4***: Use six sub keys for the first round, and the first two for the second round.
***Step5***: Rotate key 25 bits to the left and again divide into eight sub keys
***Step6***: Use first four in round 2; the last four round 3.
***Step7***: Rotate the key another 25 bits to the left for the next eight sub keys, and so on until the termination of the algorithm
***Step8***: End
In consumer's end the encrypted message is decrypted in the above-mentioned manner

# 5. Proposed Object Oriented Modelling (OOM)

Proposed model used four classes namely, CommonFields, SecurityInterface, Sender and Receiver. Relationship among different classes is shown in figure 2 with the help of class diagram. The concept contains following classes which are further discussed below:

## 5.1. CommonFields

Data Members:

        msg, key, name, age, f_votercard

Member methods:

        setall(), getage(), getkey(), getmsg()
        getf_votercard(), setf_votercard()
        getbasicdata(), showbasicdata()

Role of importance member methods are as follows:
setall() – sets all the data member to default values.
getage() – input the age of the consumer.
getkey() – receives the secret key using which the message will be encrypted and decrypted respectively.
getmsg() – receives the message that is to be encrypted by the Government.
getbasicdata() – receives all the basic information of the consumer.
showbasicdata() – displays all the basic information of the consumer.
setf_votercard() – grants the voting right to the consumer.
getf_votercard() – checks the status of voting right of the consumer.

## 5.2 SecurityInterface

Member methods:

        showinitialstatus(), sendforenc(), mul()
        mulInv(), ideaExpandKey(), ideaInvertKey(), ideaCipher()
        ideaCfbReinit(), ideaCfbinit()
        ideaCfbDestroy(), ideaCfbEncrypt()
        ideaCfbDecrypt(), ideaRandInit()
        ideaRandByte()
        Encryption, Decryption()

Roles of important member methods are as follows:

showinitialstatus() – displays default information of the consumer.
Encryption() – performs the encryption of the message using IDEA algorithm at the Government end.
Decryption() – performs the decryption of the encrypted message using IDEA algorithm at the consumer end.
ideaExpandKey() – Performs the expansion of the  128-bit user key to a working encryption key
ideaInvertKey() – Produces the corresponding key for decryption
mulInv()- Computes the multiplicative inverse of x, modulo 65537, using Euclid's algorithm.

**5.3 Sender**

Member methods:

      setvotercardflag()

Role of important member method is as follows:

setvotercardflag() – method related to granting the voting right of the consumer at the Government end.

**5.4  Receiver**

Member methods:

      checkoverallstatus()

Role of important member method is as follows:

checkoverallstatus() – method to check all the facilities available to the consumer after decryption of the encrypted message is done.

The description of CommonFields class is as follows:

This class contains the basic attributes of an individual which can be further enhanced. "Message" and "key" are used for implementation of IDEA algorithm. The length of "msg" is 64 bit and "key" is 128 bit. The member functions of class CommonFields are basically meant for input and output operation of various attributes of an individual.

In security interface the class does not contain any data members of its own, but access the those of CommonFields class through member methods of CommonFields class. This is possible because SecurityInterface class is publicly inherited from CommonFields class. This class contains various methods which are necessary for implementation of IDEA algorithm. The main two methods which performs the encryption and decryption operations are encryption () and decryption ().

The sender class represents the Government since the authors had implemented IDEA algorithm over G2C model of e-Governance. setvotercardflag() is the only member method of this class. The task of this method is to grant the voting right of the individual. With the enhancement of this concept the member methods of this class will go on increasing.The reciver class contains only one member method whose task is to display all the attributes of an individual when required.

The above-mentioned ideas are further graphically described below using sequence diagram and the class diagram [16].

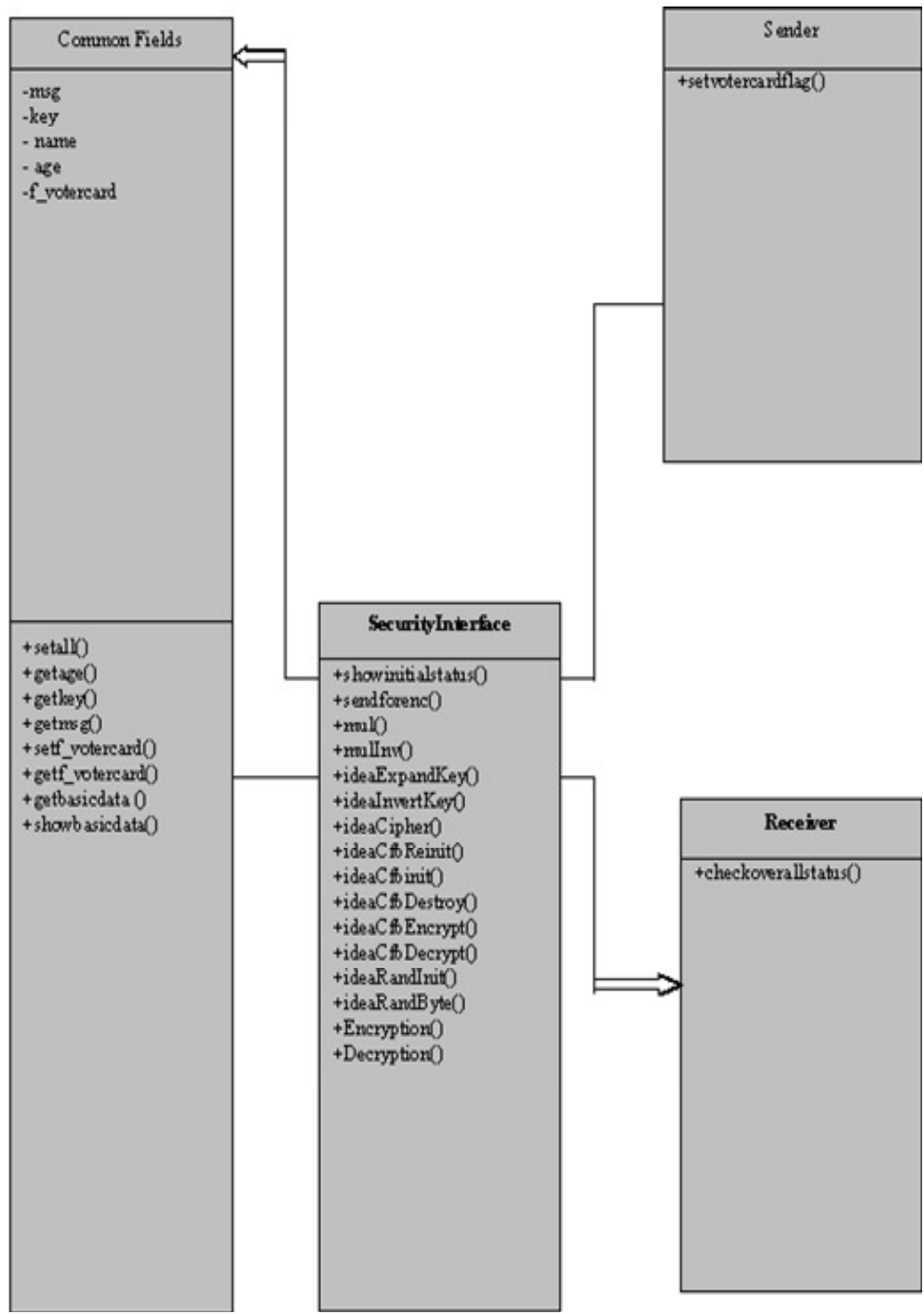

Figure 2.        Class Diagram

The sequence diagram shows the entire process execution steps. Once the birth certificate is issued for an individual, the consumer detail will go to the government. Using IDEA (International data encryption algorithm), the information will be encrypted, so that government can protect individual's information and along with can restrict the unauthorized users or intruder. Because information encryption is the most important issue in this context. This will ensure the citizenship of the country. Then a unique ID will be generated. This unique ID will be transmitted to the consumer. Based on unique ID, the information will be again decrypted to know the various services those are available on the instrument. The encrypted information can only be decrypted using IDEA (International data encryption algorithm).When a death certificate is issued to the consumer, the information will also be forwarded to the government, which will in turn stop the services those were available on the instrument.

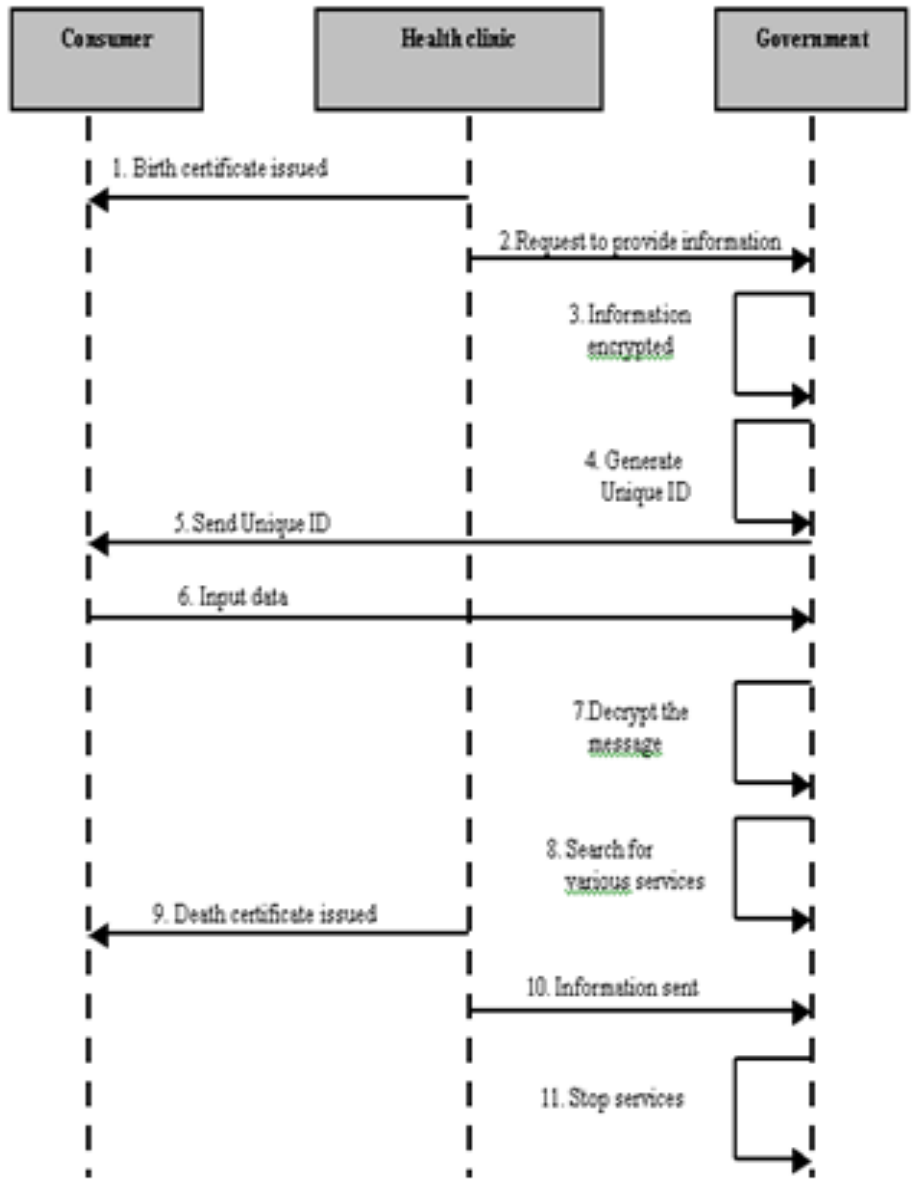

Figure 3.        Sequence Diagram

The sample output of the simulation of the proposed technique is given in figure 4.

Figure 4.        Sample Output

## 6. FUTURE SCOPE & CONCLUSION

This proposed Multipurpose Electronic Card (MEC) may be incorporated successfully by wrapping IDEA algorithm with an object oriented model of G2C which would be efficient, robust and secured. Through object orientation we have developed a new version of IDEA algorithm which may be used for e-Governance security. In this version of IDEA algorithm genetic algorithm has played a great role for key generation purpose. This genetic technique enhanced the security of the IDEA algorithm by providing random key for each session and the quality of the key is maintained by fitness function of the algorithm. But there is also scope for future enhancements. Further research work can be done to increase the acceptability of this Multipurpose Electronic Card (MEC) by implementing biometric authentication of the card's owner.

## REFERENCES

[1]     Sur C., Roy A. and Banik S. (2010), "A Study of the State of E-Governance in India" – Proceedings of the NACCS 2010 pp. a-h.

[2]     E-governance Website: *http://en.wikipedia.org/wiki/E-governance*, Date of access - 10[th] Oct, 2010.

[3]     Department of Information Technology, Ministry of Communications and Information Technology, Government of India, *National E-Governance Plan* Website: *http://india.gov.in/govt/national_egov_plan.php*, Date of access - 10[th] Oct ,2010.

[4]     Echoupal, Website: *http://en.wikipedia.org/wiki/Echoupal*, Date of access - 10[th] Oct, 2010.

[5]     TINXSYS, Website: *www.tinxsys.com*, Date of access - 10[th] Oct, 2010.

[6]     E-Governance Initiatives – India [Online].  Website: *www.mapit.gov.in/compendium.pdf*, Date of access - 10[th] Oct, 2010.

[7]     India Development Gateway, Website: *http://www.indg.in/e-governance/ict-initiatives-in-states-uts/e-governance/ict-initiatives-in-states-uts/ict-initiatives-in-states-uts*, Date of access – 10[th] Oct, 2010.

[8]     Banerjee S. and Karforma S. (2008), "A Prototype Design for DRM based Credit Card transaction in E-commerce" – ACM Ubiquity Vol.9 Issue 18 pp. 1-1.

[9]     Lai X., Massey J. and Murphy S. (1991), "Advances in Cryptology" - EUROCRYPT '91, *Lecture Notes in Computer Science*, pp. 17-38. Springer-Verlag, New York, NY.

[10]    Schneier B. (1996), "*Applied Cryptography*", Wiley, New York, NY.

[11]    Government to Consumer Model, Website: *http://en.wikipedia.org/wiki/G2C* , Date of access - 10th Oct, 2010.

[12]    White, Jay D. (2007), "Managing Information in the Public Sector." M.E. Sharpe. New York.

[13]    Soukaena H. Hashem, Mohammad A. AL-Hamami, Alaa H. AL-Hamami, "Developing a Block-Cipher-Key Generator Using Philosophy of Data Fusion Technique " *Journal of Emerging Trends in Computing and Information Sciences"*, Volume 2 No.5, MAY 2011 ISSN 2079-8407.

[14]    Gove Nitinkumar Rajendra, Bedi Rajneesh kaur , "A New Approach for Data Encryption Using Genetic Algorithms and Brain Mu Waves", International Journal Of Scientific And Engineering Research Volume 2, Issue 5, May-2011 1, ISSN 2229-5518.

[15]    C code implementation of IDEA algorithm Website: *http://www.koders.com/c/ fid43C02E2565B7947584D23C36A6C32E18E198E06C.aspx?s=des,*Date of access - 10th Oct, 2010.

[16]    Bézivin J., Muller P-A. (1998), "The unified modeling language: UML '98: beyond the notation" First International Workshop, Mulhouse, France, June 1998, SPRINGER.

[17]    Mandal  J. K., Sarkar Arindam, "Neural Session Key based Traingularized Encryption for Online Wireless Communication (NSKTE)", 2nd NATIONAL CONFERENCE ON COMPUTING AND SYSTEMS, (NaCCS 2012),  March 15-16, 2012, Department of Computer Science, The University of Burdwan, Golapbag North, Burdwan –713104, West Bengal, India.

[18]    Mandal  J. K., Sarkar Arindam, "Neural Weight Session Key based Encryption for Online Wireless Communication (NWSKE)", Research and Higher Education in Computer Science and Information Technology, RHECSIT- 2012 ,  February 21-22, 2012, Department of Computer Science, Sammilani Mahavidyalaya, Kolkata , West Bengal, India.

[19]    Mandal  J. K., Sarkar Arindam, "An Adaptive Genetic  Key Based  Neural Encryption For Online Wireless  Communication  (AGKNE)",  INTERNATIONAL CONFERENCE ON RECENT TRENDS IN INFORMATION SYSTEM (RETIS 2011) BY IEEE, held on 21-23 December 2011, , Jadavpur University, Kolkata, India. ISBN 978-1-4577-0791-9.

[20]    Mandal J. K., Sarkar Arindam, "An Adaptive Neural Network Guided Secret Key based Encryption through Recursive Positional Modulo-2 Substitution for Online Wireless Communication (ANNRPMS)", INTERNATIONAL CONFERENCE ON RECENT TRENDS IN INFORMATION TECHNOLOGY (ICRTIT 2011) BY IEEE, held on 3-5 June 2011, Madras Institute of Technology, Anna University, Chennai, Tamil Nadu, India. ISBN 978-1-4577-0590-8/11

[21]    Mandal J. K., Sarkar Arindam, "An Adaptive Neural Network Guided Random Block Length Based Cryptosystem (ANNRBLC)", ID: 186, Special Session: Security Protection Mechanism in Wireless Sensor Networks, IEEE 2nd INTERNATIONAL CONFERENCE ON WIRELESS COMMUNICATIONS, VEHICULAR TECHNOLOGY, INFORMATION THEORY AND AEROSPACE & ELECTRONIC SYSTEM TECHNOLOGY" (WIRELESS VITAE 2011) BY IEEE SOCIETIES, held on February 28, 2011- March 03, 2011, Chennai, Tamil Nadu, India. ISBN 978-87-92329-61-5,p. 57,2011.

[22]    Mandal  J. K., Sarkar Arindam, "Neural Network Guided Secret Key based Encryption through Cascading Chaining of Recursive Positional Substitution of Prime Non-Prime (NNSKECC)", INTERNATIONAL  CONFERENCE  ON  COMPUTING  AND  SYSTEMS,  (ICCS 2010), November 19, 2010– November 20, 2010,Department of Computer Science, The University of Burdwan, Golapbag North, Burdwan –713104, West Bengal, India. ISBN 93-80813-01-5.

**Arindam Sarkar**

Presently Research Scholar of University of Kalyani, M.Tech ((CSE), University of Kalyani, University First Class First Rank Holder), MCA (VISVA BHARATI, Santiniketan, University First Class First Rank Holder)

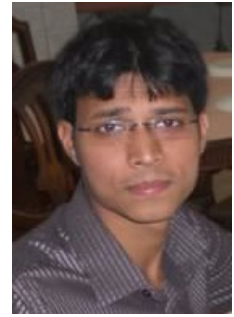

**Sunil Karforma**

BE (Computer Science and Engineering) and ME (Computer Science and Engineering) from Jadavpur University. He has completed Ph. D. in the field of Cryptography. He is presently holding the post of Reader and the Head of the Department in the Department of Computer Science, The University of Burdwan. Network security and e-commerce are his field of interest in research area. He has published approximately 16 research papers in reputed National and International journals and proceedings.

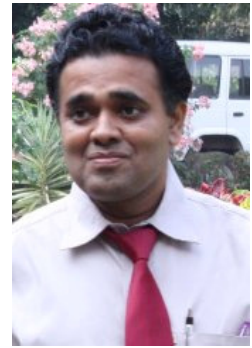

**Jyotsna Kumar Mandal**

M. Tech.(Computer Science, University of Calcutta), Ph.D.(Engg., Jadavpur University) in the field of Data Compression and Error Correction Techniques, Professor in Computer Science and Engineering, University of Kalyani, India. Life Member of Computer Society of India since 1992 and life member of cryptology Research Society of India. Dean Faculty of Engineering, Technology & Management, working in the field of Network Security, Steganography, Remote Sensing & GIS Application, Image Processing. 25 years of teaching and research experiences. Eight Scholars awarded Ph.D. and 8 are pursuing.  Total number of publications 187.

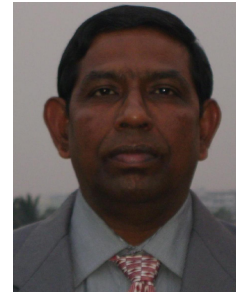